\documentclass[12pt]{article}
\usepackage{times}
\usepackage{geometry}
\usepackage[utf8]{inputenc}
\usepackage{enumitem,amssymb}
\usepackage{ragged2e}
\newlist{thematic}{itemize}{8}
\setlist[thematic]{label=$\square$}
\usepackage{pifont}
\newcommand{\cmark}{\ding{51}}%
\newcommand{\done}{\rlap{$\square$}{\raisebox{2pt}{\large\hspace{1pt}\cmark}}%
\hspace{-2.5pt}}

\usepackage{fancyhdr}
\usepackage{natbib}
\setcitestyle{aysep={}}
\usepackage{graphics,graphicx}
\usepackage{color}
\usepackage{xcolor}
\usepackage{lineno}
\usepackage{wasysym}
\usepackage{wrapfig}

\newcommand{\msol}{M$_\odot$}

\def\apj{\rm ApJ}
\def\apjl{\rm ApJL}

\def\mnras{\rm MNRAS}

\begin{document}
\begin{flushleft}
\huge
Astro2020 Science White Paper \linebreak

A Summary of Multimessenger Science with Galactic Binaries \linebreak
\normalsize

\noindent \textbf{Thematic Areas:} \hspace*{60pt} $\square$ Planetary Systems \hspace*{10pt} $\square$ Star and Planet Formation \hspace*{20pt}\linebreak
$\done$ Formation and Evolution of Compact Objects \hspace*{31pt} $\square$ Cosmology and Fundamental Physics \linebreak
  $\done$  Stars and Stellar Evolution \hspace*{1pt} $\square$ Resolved Stellar Populations and their Environments \hspace*{40pt} \linebreak
  $\square$    Galaxy Evolution   \hspace*{45pt} $\done$             Multi-Messenger Astronomy and Astrophysics \hspace*{65pt} \linebreak
  
\textbf{Principal Author:}

Name: Thomas Kupfer
 \linebreak						
Institution: Kavli Institute for Theoretical Physics / UC Santa Barbara
 \linebreak
Email: tkupfer@ucsb.edu
 \linebreak
Phone: +1-805-893-6326
 \linebreak
 
\textbf{Co-authors:} Mukremin Kilic (University of Oklahoma), Tom Maccarone (Texas Tech), Eric Burns (NASA Goddard), Chris L. Fryer (Los Alamos National Laboratory), Colleen A. Wilson-Hodge (NASA Marshall)
  \linebreak

\end{flushleft}

\textbf{Abstract:} Galactic binaries with orbital periods less than $\approx$1\,hr are strong gravitational wave sources in the mHz regime, ideal for the {\it Laser Interferometer Space Antenna} (\emph{LISA}). In fact, theory predicts that \emph{LISA} will resolve tens of thousands of Galactic binaries individually with a large fraction being bright enough for electromagnetic observations. This opens up a new window where we can study a statistical sample of compact Galactic binaries in both, the electromagnetic as well the gravitational wavebands. Using multi-messenger observations we can measure tidal effects in detached double WD systems, which strongly impact the outcome of WD mergers. For accreting WDs as well as NS binaries, multi-messenger observations give us the possibility to study the angular momentum transport due to mass transfer. In this white paper we present an overview of the opportunities for research on Galactic binaries using multi-messenger observations and summarize some recommendations for the 2020 time-frame.

			



\pagebreak

\section{Introduction}
Ultracompact binaries (UCBs) are an exotic class of (semi-)detached binary stars with orbital periods as short as a few minutes and separations of the order of the Earth-Moon distance. This is only possible if {\it both} components are evolved objects consisting of a neutron star (NS), white dwarf (WD) or in rare cases black hole (BH) primary and a Helium-star/WD/NS secondary. These systems are the dominant Galactic sources in the \emph{LISA} gravitational wave band and are crucial to our understanding of compact binary evolution and offer pathways towards Type Ia supernovae.

The formation mechanism for UCBs involves two common envelope phases that produce a detached double WD binary system, which is brought into contact at a period between 3-10 minutes through gravitational wave emission \citep{podsi03}. In rare cases, if either one of the stars has an initial mass of $M\geq$8-10\,\msol\, the star will explode as a supernova after the first or second phase of mass transfer and end up as a NS or BH rather than a WD. This results in either a WD + NS/BH  or a He-star + NS/BH system.

If the system survives the onset of mass-transfer [this process and the stability of mass transfer are not understood at all, \citealt{nelemans01, marsh04}], and accretes stably onto a WD or a NS, the result is an AM\,CVn or an ultra-compact X-ray binary (UCXB), respectively. Stable mass transfer in these systems leads to longer periods as the orbits widen. If the system does not survive the onset of mass-transfer, both WDs merge and either form an RCrB star or explode as a supernova Ia if the total mass is above the Chandrasekhar limit.

\emph{LISA} is a space-based gravitational detector sensitive to lower frequencies than \emph{LIGO}. UCBs are strong gravitational wave (GW) sources and will dominate the population of gravitational wave emitters in the \emph{LISA} band. Systems with orbital periods $<20$min will be the strongest Galactic \emph{LISA} sources and will be detected by \emph{LISA} within weeks after its operation begins. These `verification binaries' are crucial in facilitating the functional tests of the instrument and maximize \emph{LISA}'s scientific output. Galactic binaries with periods $\lesssim$2\,hrs are predicted to be so numerous that individual detections are limited by confusion with other binaries yielding a stochastic foreground or confusion signal.

\begin{wrapfigure}{r}{0.44\textwidth}
\includegraphics[width=0.42\textwidth]{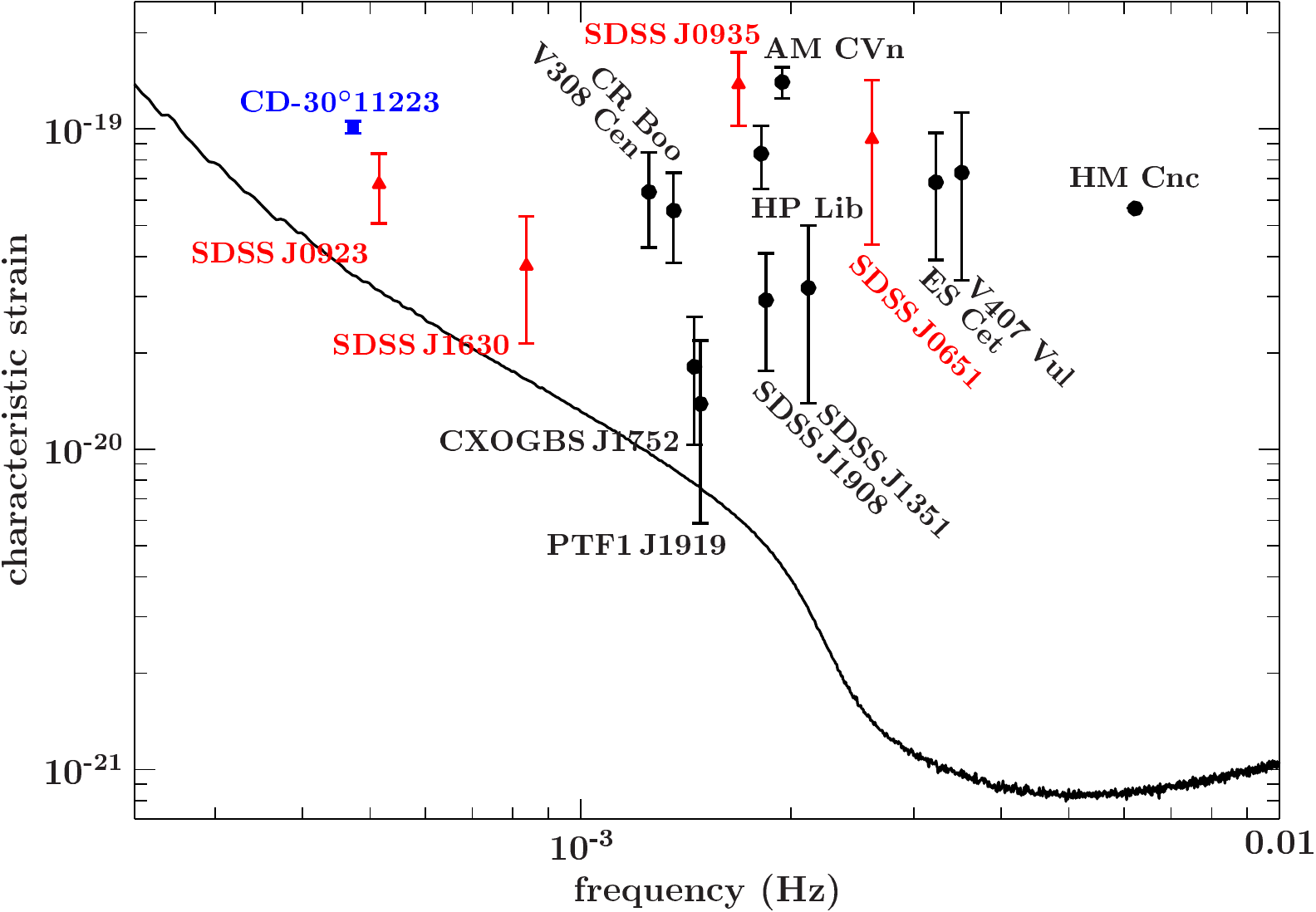}
\caption{\small{Sensitivity limits of \emph{LISA} shown with the known verification systems and the evolutionary paths of different UCBs. Black circles are AM\,CVn systems, red triangles correspond to detached double WDs and the blue square is an ultracompact He-star + WD \citep{kupfer18}.}}
\label{fig:gw_plot}
\end{wrapfigure}

Binary population studies predict that we will be able to detect in GW and electromagnetic (EM) several thousand detached and semi-detached double WDs as well as a few tens of NS or black hole binaries with a population strongly peaking towards the Galactic Plane/Bulge (e.g. \citealt{nelemans04}). Observationally, we know of only about a \underline{dozen} although hundreds to thousands are predicted to be detectable in our Galaxy (Fig.\,\ref{fig:gw_plot}; \citealt{kupfer18}). Additionally, the known verification binary sample is strongly biased and incomplete, with semi-detached AM\,CVn type binaries strongly overrepresented. Here, we present an overview of potential science for Galactic binaries using multi-messenger observations and finish with recommendations for the 2020s time-frame. 

\section{Astrophysics using combined EM+GW observations}

WD binaries will be among the best laboratories for understanding the formation of compact objects, dynamical interactions and tides in binaries, and  Type Ia supernovae. Combined EM/GW observations yield more robust masses, radii, orbital separations, and inclination angles than either   can achieve on its own. \citet{shah12} found a strong correlation between GW amplitude and inclination, and demonstrated that EM constraints on the inclination can improve the GW amplitude measurements by up to a factor of six. In addition, knowing the sky position and inclination can reduce the uncertainty in amplitude by up to a factor of 60 \citep{shah13}. Similarly, using the chirp mass obtained from GW observations and the mass ratio from spectroscopic radial velocity measurements allows an independent measurement of the masses of both components to exquisite precision. This enables a direct comparison between the rate of orbital decay observed in GW or EM (for eclipsing and/or tidally distorted systems) and predicted from GR. 

Measuring the effects of tides in binaries, which are predicted to contribute up to 10\% of the orbital decay, is almost impossible from GW data alone \citep{shah14}, but the EM data on distance constrains the uncertainty in chirp mass to 20\%, whereas adding $\dot{P}$ reduces it to 0.1\%. A GW chirp mass measurement would provide the first detection of tidal heating in a merging pair of WDs from the deviations in predicted $\dot P$.

\subsection{Detached double white dwarfs}
There are now more than 100 short period detached double WDs known, thanks to large scale surveys like the SN Ia Progenitor Survey \citep{napiwotzki04} and the Extremely Low-Mass (ELM) WD survey \citep{kilic10,brown10}. The current sample of short period binaries is dominated by ELM WDs with orbital periods ranging from 12 min to $\sim1$ day \citep{brown11,brown16}.  

EM observations of detached double WDs provide precise sky positions, mass ratios (especially for the double-lined spectroscopic binaries), inclinations \citep[][for the $\sim10$ eclipsing or tidally distorted systems currently known]{hermes14}, and the rate of orbital period decay. Looking forward to EM+GW multi-messenger astrophysics, we highlight J0651, the 12 min orbital period system, as an example of what can be accomplished. \citet{hermes12} provided eclipse time measurements of J0651 and found that the eclipses occurred $6.1 \pm 0.6$\,s earlier than expected over a 1 year period if the orbital period stayed the same. This corresponds to a rate of orbital decay of $\dot P = (-9.8\pm 2.8)\times 10^{-12}$\,s\,s$^{-1}$. The $\dot P$ measurement is now much more precise (J.J. Hermes, private communication). However, we do not know the masses of the WDs in this system well enough from optical observations to see if J0651's $\dot{P}$ differs from the GR predictions. GW observations can solve that. Tidal theory predicts a 10\% deviation from GR if the WDs are tidally heating up \citep{benacquista11,piro11,fuller12,fuller13}.  Which means that combining EM+GW observations will allow a measurement of the amount of tidal heating in a merging pair of WDs for the first time.

Combined GW+EM observations of the Galactic WD population will help to solve another major problem in astrophysics; the SNe\,Ia progenitor problem. Short period double WDs with a total mass near the Chandrasekhar limit are one of the proposed progenitor channels for SNe\,Ia, yet we have no convincing progenitor system identified to date. \citet{rebassa19} studied the probability of finding double WD progenitors of SNe\,Ia using a binary population synthesis approach, and found that the chance of identifying such progenitors purely in EM data is $\sim10^{-5}$. These include both double-lined spectroscopic binaries and the eclipsing systems. Even with next generation of 30m class telescopes, the probability for detection only goes up by a factor of $\sim10$. \citet{korol17} predicts that \emph{LISA}, will individually resolve $\sim25,000$ detached double WD systems including the most massive systems. EM follow-up observations in combination with GW measurements will allow us to measure masses of individual systems and find and characterize the population of double degenerate SN\,Ia progenitors.  

\subsection{Semi-detached double white dwarfs (AM\,CVn binaries)}
In their evolution, double WD binaries do not always remain detached. If the system survives the onset of mass-transfer and accretes stably onto a WD a so-called ultracompact AM\,CVn binary is formed. AM\,CVn systems evolve towards longer periods as the orbit widens. Their number and properties provide an observational anchor for our understanding of WD accretion physics and mass transfer stability. The largest uncertainties in predictions of the final fate of a double WD binaries (merger versus stable mass transfer) and hence of SN\,Ia rates is the treatment of the onset of mass transfer when the larger WD fills its Roche Lobe and starts to accrete onto its companion. So far, the number of known AM\,CVn systems in the Galaxy is significantly smaller than predicted by binary population studies which challenges our theoretical understanding of their formation and evolution \citep{carter13,ramsay18}. Double WD systems with mass ratios $q=M_2/M_1<2/3$, $M_1$ being the mass of the accretor, are commonly assumed to form AM\,CVn type systems. Systems with larger mass ratios will merge. \citet{marsh04} and \citet{gokhale07} studied the effect of coupling of the accretor's and donor's spin to the orbit. They found that a strong coupling and therefore a strong feed back of angular momentum to the orbit can destabilize systems with mass ratios lower than $2/3$. This in turn has strong impact on the individual population. Most recently, \citet{shen15} proposed that even accreting double WD binaries with extreme mass ratios will merge due to classical nova-like outbursts on the accretor. Dynamical friction within the expanding nova shell causes the binary separation to shrink and the donor to dramatically overfill its Roche lobe, resulting in highly super-Eddington mass transfer rates that lead to a merger. Combining EM and GW observations will provide for the first time answers on the stability of mass-transfer as well as accretion disc physics in accreting double WDs.


The large statistical sample of accreting and non-accreting double WDs detected by \emph{LISA} and confirmed by EM will provide us with space densities and system properties of each group individually. This in turn gives a direct measure on how many double WDs avoid the merger as well as what are the system properties of the surviving AM\,CVn binaries. Comparing space densities derived from GW data and system properties (like e.g. mass ratios) derived from EM observations of non-accreting and accreting double WDs will give us unique insights on the physics when both WDs get into contact.

The angular momentum transport in AM\,CVn type binaries is governed by two competing processes; gravitational wave radiation shrinks the orbit while mass transfer widens the orbit. EM observations can provide masses, orbital periods, sky positions and, sometimes distances, while GW measurements can provide sky positions, orbital periods and their derivatives ($\dot{P}$) and in some cases chirp masses. With these quantities in hand we can disentangle the contribution from gravitational wave radiation and mass transfer from the overall $\dot{P}$ and study, for the first time, the transport of angular momentum in accreting double WDs on a statistically significant number of systems. The strength of angular momentum transport is intimately related to how much mass is being accreted in the system. A comparison of the accretion rate with the X-ray and UV luminosities and spectra will allow a deeper insight into the radiative properties of matter.

\subsection{Neutron star - white dwarf binaries \& black hole-white dwarf binaries}

Three important classes of NS-WD binaries are -- those with mass-transfer, UCXBs; those which have not yet evolved into Roche lobe overflow and have not yet shown mass transfer, the "pre-UCXBs"; and those which are the end stages of recycling of millisecond pulsars, which may also re-start mass transfer at a later time.  

One of the core uncertainties in understanding the evolution of pulsars is the determination of just how much mass is accreted to spin the pulsars up -- low mass X-ray binaries are often seen with donor stars of $\approx 1 M_\odot$, but millisecond pulsars only rarely show masses more than about 1.5 $M_\odot$, suggesting that only rarely the whole donor star is accreted. With \emph{LISA}, one can expect to develop substantial-sized samples of both the pre-ultracompact X-ray binaries and the systems that are still accreting at low rates. Furthermore, precise masses for all components will be possible in these systems. This will then allow empirical estimates of both the amount of mass lost by the WDs and the amount of mass gained by the NSs, allowing a clear test of how conservative the mass transfer is in these systems. Over the past decade, some evidence has started to emerge for BH-WD binaries in globular clusters \citep{maccarone07,bahramian2017}, and the populations of these sources, too, can be investigated with \emph{LISA}. A goal for EM follow-up of both classes of sources (WD+NS and WD+BH systems) is the fraction of helium WDs relative to carbon-oxygen WDs in these systems, which requires spectroscopy on 8-10 meter class optical telescopes.

A second important goal that can be achieved by the use of WD-NS binaries discovered by \emph{LISA} is to understand the opening angles for pulsar beams at different energy bands.  \citet{tauris18} predicts that a large fraction of ultracompact binaries should be formed after going through phases of existence first as normal low mass X-ray binaries, then as detached WD-pulsar binaries. During much of the latter phase, these sources should be within the \emph{LISA} bandpass. Here, the masses of the NSs can be measured from LISA. Furthermore, crucially, this set of objects will provide a {\it gravitationally-selected} sample of recycled NSs. With a set of facilities optimized for detection of pulsations in radio, X-ray and gamma-rays, it will then be possible to determine which objects are detected as pulsars in which wavelengths. At the present time, it appears likely that pulsars have wider opening angles at higher energies than at radio \citep{ravi2010}, but modelling of poorly understood selection effects is required to interpret all electromagnetically selected samples. A 2-$m^2$ X-ray mission with a background rate of 10 $\mu$Crab would be able to detect the pulsations for pulsars with X-ray luminosities of $10^{31}$ erg/sec and pulse fractions of 10\% at distances of about 6 kpc in reasonable exposure times.

Finally, ultracompact WD-NS and WD-BH binaries are being the proposed progenitors of energetic explosions. As the orbit tightens, the WD is tidally disrupted by its NS or BH companion, forming a debris disk around the compact companion. If magnetic fields develop, this merger can produce a ultra-long duration gamma-ray burst \citep{fryer1999}. Alternatively, explosive burning in this disk can power a supernova-like outburst, and these outbursts have been proposed to explain Ca-rich supernovae \citep{metzger2012}. 

\section{Prospects of Multi-messenger observations of Galactic binaries in the 2020s}

Binary population studies predict \emph{LISA} will resolve a few tens of thousands of Galactic binaries. Systems with periods of a few minutes will be seen throughout the Galaxy some of them already within the first few weeks of \emph{LISA} observations. Thanks to their high signal-to-noise ratio they will be well localized, thus making them excellent candidates for multi-messenger observations. 

The known population of verification binaries is biased and in-homogeneous. However, this will likely change in the upcoming years. Compact binaries can be detected through time-series photometry as their lightcurves show variations on timescales of the orbital period (e.g. due to eclipses or tidal deformation of the components), and also through multi-epoch spectroscopy that may reveal large radial velocity shifts between individual spectra. Therefore, photometric and spectroscopic time-domain surveys are well suited to identify Galactic \emph{LISA} sources in an homogeneous way. We are in an era with many emerging new large scale time domain surveys such as ZTF, ATLAS, LSST, \emph{Gaia}, SDSS-V, and DESI. 

\emph{Gaia} and LSST both include photometric variability surveys with cadences of about 26 and 3 days, respectively. Gaia\ is an astrometric mission, launched in December 2013, that is delivering a three-dimensional map of stars down to a G-band magnitude of 20.7 over the entire celestial sphere. In addition to Gaia's relatively low-cadence observations, on average once every 26 days, a catalog of 260,000 WD candidates came out of Gaia DR 2 \citep{fusillo19} and are being followed-up now. As of this writing LSST is under construction with a planned completion date in the early 2020's. LSST will observe the southern sky down to an r-band magnitude of ${\sim}24$, discovering millions of WDs. Although, LSST's cadence is not ideal for the discovery of new verification binaries, it is well suited to identify and study the EM counterparts of Galactic binaries detected by \emph{LISA}, which is predicted to localize several thousand ultracompact binaries to better than one square degree after a few years of observations \citep{cornish17}. LSST's limiting magnitude will allow us to detect the EM counterparts of accreting verification binaries throughout the Galaxy (if not hidden by Galactic dust) and for detached systems to 10 kpc. 

We expect that many new \emph{LISA} verification binaries will be discovered and will be complemented with surveys in other frequency bands (radio to gamma-rays). This data will be complemented with the next generation of follow-up facilities like the 30m telescope, the \emph{James Webb Space Telescope} or even \emph{Athena}. We recommend to provide support for missions across the full wavelength range, providing a "bright" future for the research on compact Galactic binaries, with dozens of additional systems ready to be studied in detail through EM+GW observations as soon as \emph{LISA} is operational.

We predict that a large number of verification binaries, observed with GW+EM observations, will open up possibilities to explore and study astrophysical phenomena which are crucial to our understanding of the universe. This includes the long-standing questions of the progenitors of supernovae Ia, the formation and evolution of compact objects in binaries and accretion physics under extreme conditions.

\end{document}